\documentclass[preprintnumbers]{revtex4}

\usepackage{graphicx}
\def\ie{{\it i.e.}}

\def\mpl{\ifmmode M_*\else $\M_*$\fi}
\def\to{\rightarrow}

\begin{document}
\bibliographystyle{revtex}

\preprint{SLAC-PUB-9053/
          P3-39}

\title{Black Hole Production Rates at the LHC: Still Large}

\author{Thomas G. Rizzo}

\email[]{rizzo@slac.stanford.edu}
\affiliation{Stanford Linear Accelerator Center, 
Stanford University, Stanford, California 94309 USA}

\date{\today}

\begin{abstract}
We examine the rates for the production of black holes(BH) at the LHC in light 
of the exponential suppression of the geometric cross section estimate 
proposed by Voloshin. We show that these rates will still be quite large over 
a reasonable range of model parameters. While BH production may not be the 
dominant process, its unique signature will ensure observability over 
conventional backgrounds. 
\end{abstract}

\maketitle

Theories with extra dimensions and a low effective Planck scale($\mpl$) offer 
the exciting possibility that the production rate of black holes(BH) somewhat 
more massive than $\mpl$ can be quite large at future colliders. For example, 
cross sections of order 100 pb at the LHC{\cite {gids}}, and even 
larger ones at the VLHC, have been advertised in the analyses presented by 
Giddings and Thomas(GT) and by Dimopoulos and Landsberg(DL). Although 
in practice 
the actual production cross section critically depends on the BH mass, 
the exact value of $\mpl$ and the number of extra dimensions, following the 
analysis of the authors in Ref.{\cite {gids}}, one 
finds very large rates over almost all of the interesting parameter space. 
These earlier analyses and discussions of the production of BH at colliders 
have been extended 
for the Snowmass proceedings by several authors{\cite {sgid}}.  

\begin{figure}[htbp]
\centerline{
\includegraphics[width=5.4cm,angle=90]{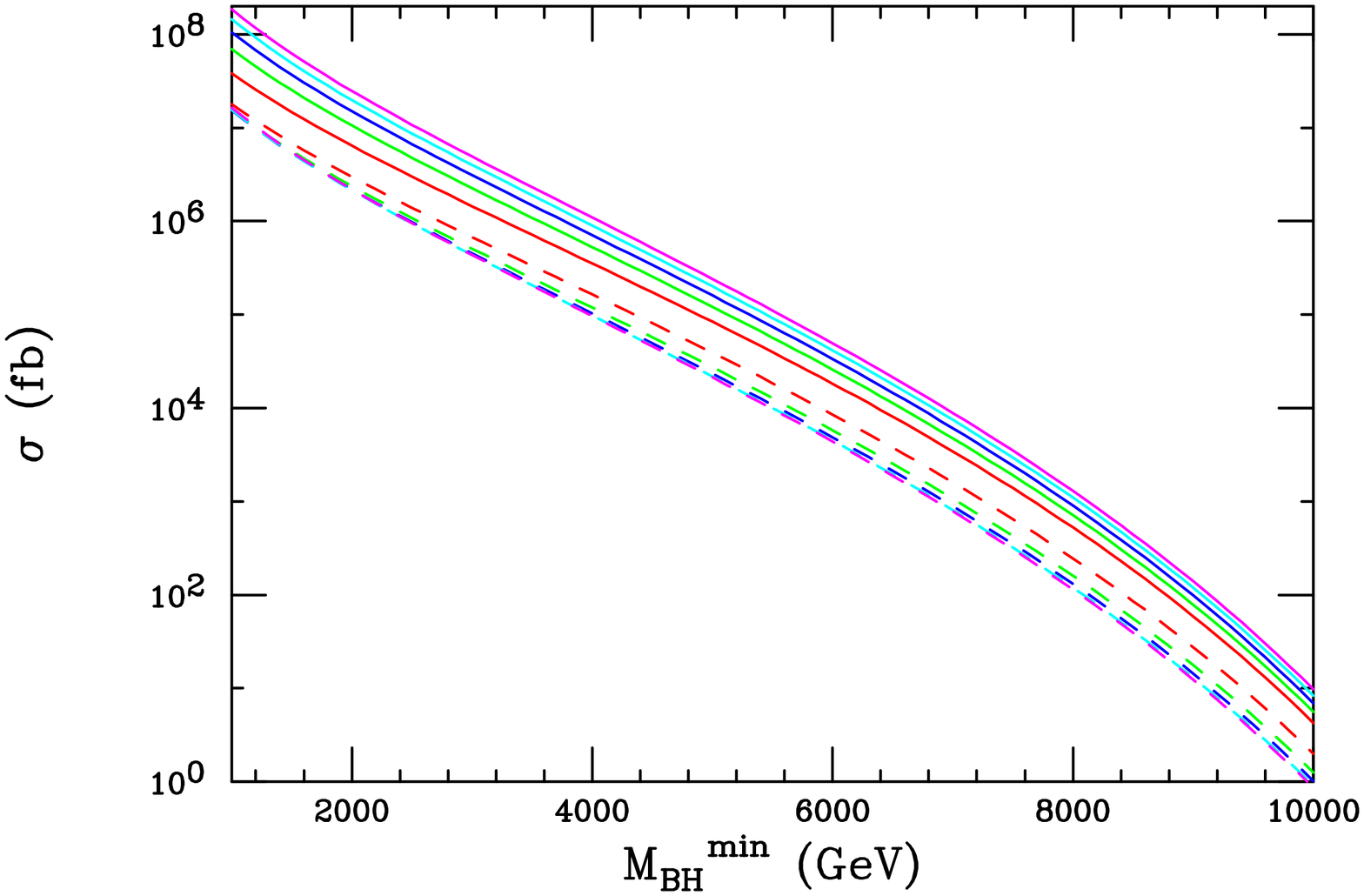}
\hspace*{5mm}
\includegraphics[width=5.4cm,angle=90]{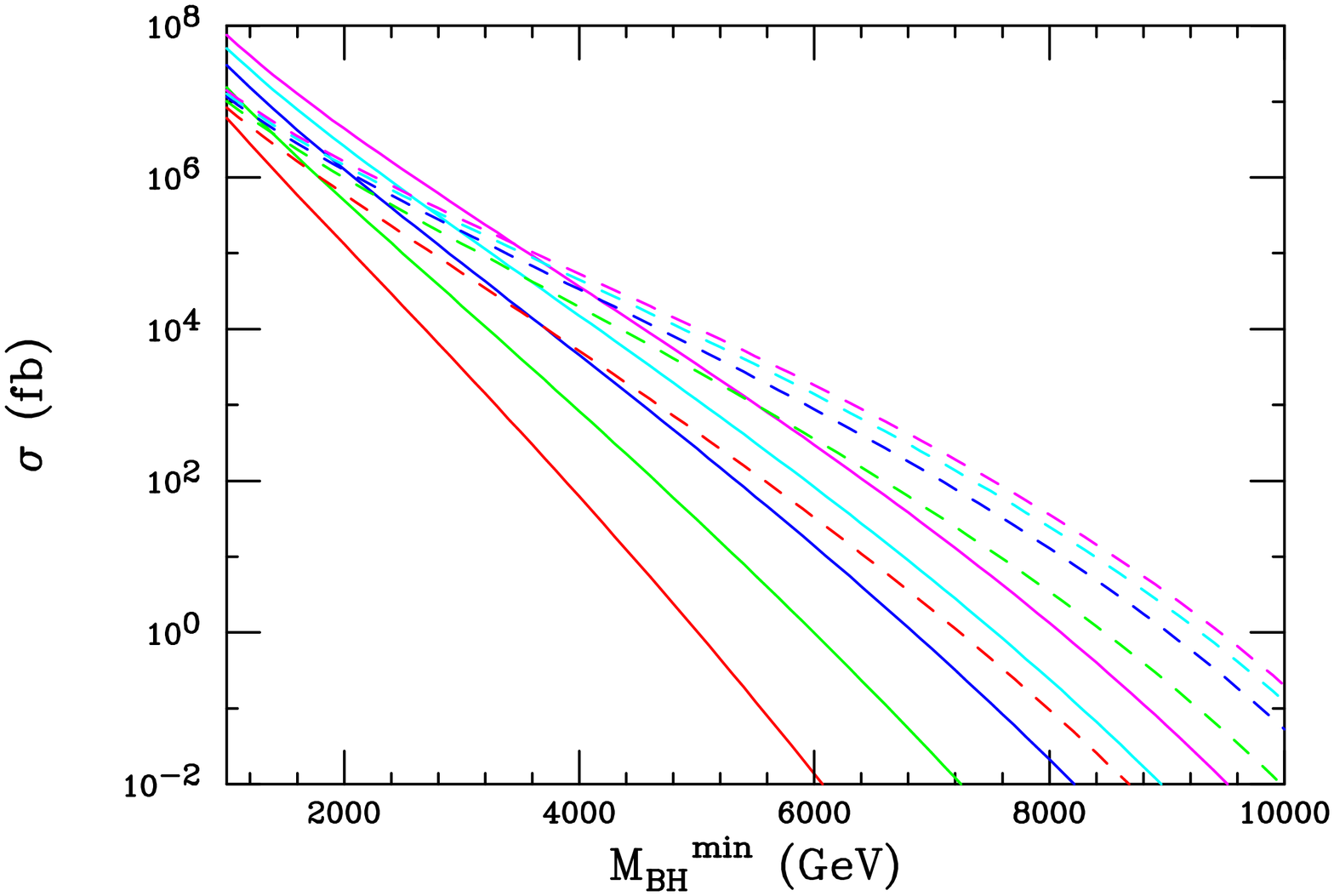}}
\vspace*{0.1cm}
\caption{(Left)Cross section for the production of BH more massive 
than $M_{BH}^{min}$ at the LHC assuming $M_* =1$ TeV for 
$\delta=2(3,4,5,6)$ extra dimensions corresponding to the 
red(green,blue,cyan,magenta) curves. 
The solid(dashed) curves are the results 
from the work by Giddings and Thomas(Dimopoulos and Landsberg). (Right)Same 
as on the left but now including the effects of the Voloshin damping factor. 
We observe that large cross sections are possible over a reasonable range of 
the model parameters. We identify $M_*$ with either $M_{GT}$ or 
$M_{DL}$ for the appropriate set of curves.}
\label{p3-33_snowholes}
\end{figure}

The basic idea behind the original collider BH papers is as follows: we 
consider the collision of two high energy Standard Model(SM) partons which are 
confined to a 3-brane, as they are in 
both the models of Arkani-Hamed, Dimopoulos and Dvali(ADD){\cite {add}} and 
Randall and Sundrum(RS){\cite {rs}}. 
In addition, we imagine that gravity is free to propagate in 
$\delta$ extra dimensions with the $4+\delta$ dimensional Planck scale 
assumed to be $\mpl \sim 1$ TeV. The curvature of the space is assumed to be 
small compared to the energy scales involved in the collision so that quantum 
gravity effects can be neglected. When these partons have a center of 
mass energy in excess of $\sim \mpl$ and the impact parameter for the 
collision is less than the 
Schwarzschild radius, $R_S$, associated with this center of mass energy, a 
$4+\delta$-dimensional BH is formed with reasonably high efficiency. 
The subprocess cross section for the production of a non-spinning BH 
is thus essentially geometric for {\it each} pair of partons: 
\begin{equation}
\hat \sigma \simeq \pi R_S^2\,;
\end{equation}
where we note that $R_S$ scales as
\begin{equation} 
R_S \sim \Big[{M_{BH}\over {\mpl^{~2+\delta}}}\Big]^{1\over {1+\delta}}\,,
\end{equation}
apart from an overall $\delta$- and {\it author-dependent} numerical factor. 
This is due to the different expressions for the $4+\delta$ dimensional 
Schwarzschild radius used by the two sets of original authors GT and DL. 
Explicitly there are two different relationships employed 
between the $4+\delta$-dimensional Planck masses, $M_{GT,DL}$, and 
the associated Newton's constant, $G_{4+\delta}$: 
$M_{DL}^{2+\delta}=G_{4+\delta}^{-1}$, while 
$M_{GT}^{2+\delta}=(2\pi)^\delta /4\pi G_{4+\delta}$. 
Depending on how the input 
parameters are chosen, this numerical 
factor can turn out to be relatively important since it 
leads to a very different $\delta$ dependence for the BH production cross 
section in the two cases. In the DL case the 
$\delta$-dependence of the numerical coefficient is rather weak whereas is it 
somewhat stronger in the GT analysis. For the same input value of 
$M_{BH}$ one finds the ratio of the cross sections obtained by 
the two sets of authors to be 
\begin{equation}
{\hat \sigma_{GT}\over {\hat \sigma_{DL}}}=\Big[{(2\pi)^\delta \over {(4\pi)
}}\Big]^{2\over {1+\delta}}~\Big[{M_{DL}^2\over {M_{GT}^2}}\Big]^{{2+\delta}
\over {1+\delta}}\,,
\end{equation}
which is always greater than unity for $\delta \geq 2$ and grows as $\delta$ 
increases {\it if} one assumes $\mpl=M_{GT}=M_{DL}$ as an input. When the 
differences in the definitions of the Planck scale are accounted for both 
cross sections lead to the {\it same} numerical result. 

The approximate geometric subprocess cross section expression is 
claimed to hold by GT and DL when the ratio $M_{BH}/\mpl$ is ``large", \ie, 
when the system can be treated semi-classically and quantum gravitational 
effects are small; one may debate just what ``large" really means, but it 
most likely means ``at least a few". Certainly when $M_{BH}/\mpl$ is near 
unity one might expect stringy effects to become important and even the finite 
extent of the incoming partons associated with this stringy-ness would need 
to be considered. 

In order to obtain the actual cross section at 
a collider one takes the geometric parton-level result, folds in the 
appropriate parton densities and integrates 
over the relevant kinematic variables. The resulting total cross section for 
BH with masses $\geq M_{BH}^{min}$ is then given by the expression  
\begin{equation}
\sigma=\int^1_{M_{BH}^{min~2}/s} d\tau \int^1_\tau {dx\over {x}} \sum_{ab} 
f_a(x) f_b(\tau /x) ~\hat \sigma(M_{BH})\,, 
\end{equation}
where we have summed over all possible pairs of initial state partons with 
their associated densities $f_i(x)$. 

\begin{figure}[htbp]
\centerline{
\includegraphics[width=5.4cm,angle=90]{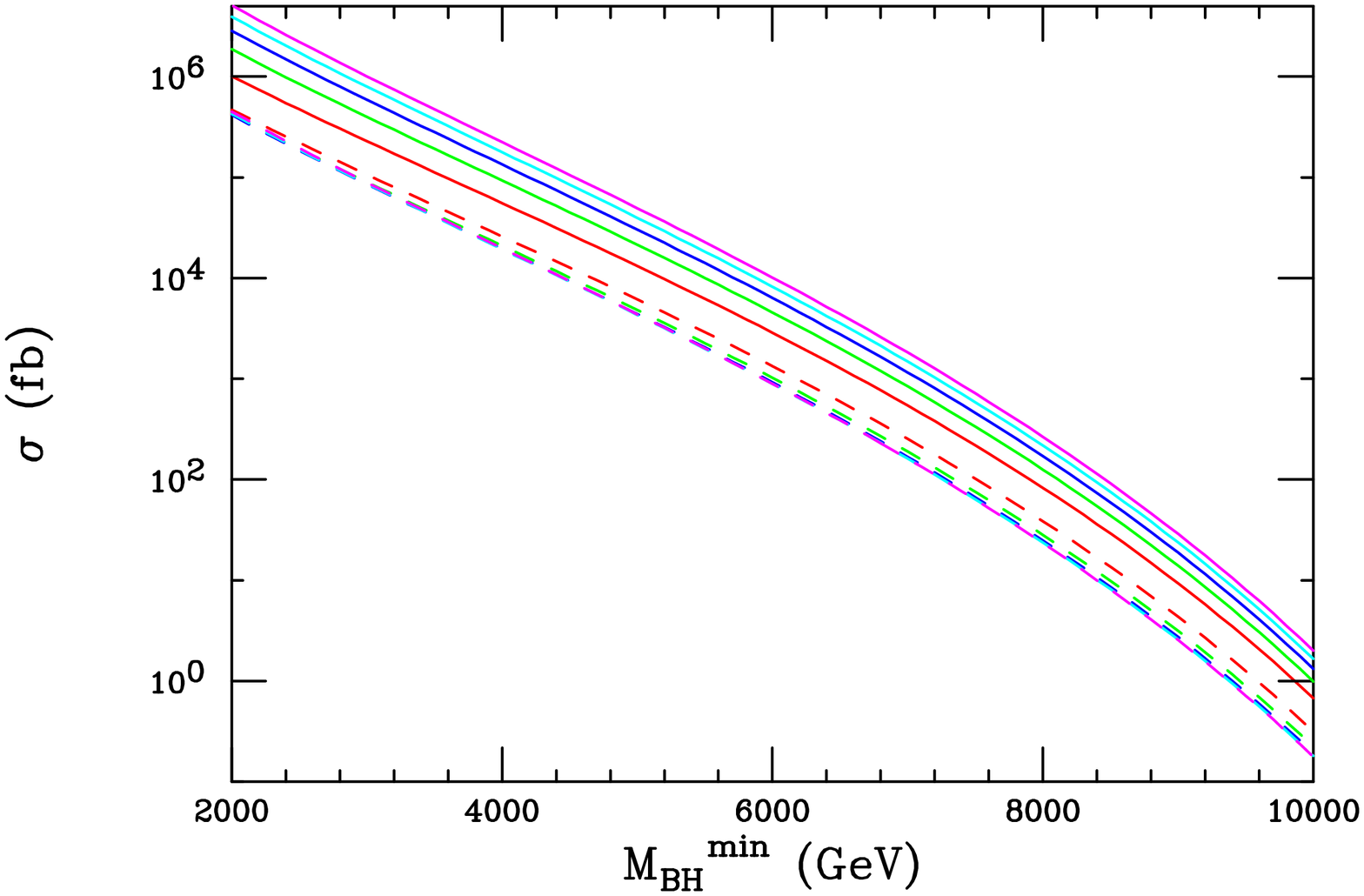}
\hspace*{5mm}
\includegraphics[width=5.4cm,angle=90]{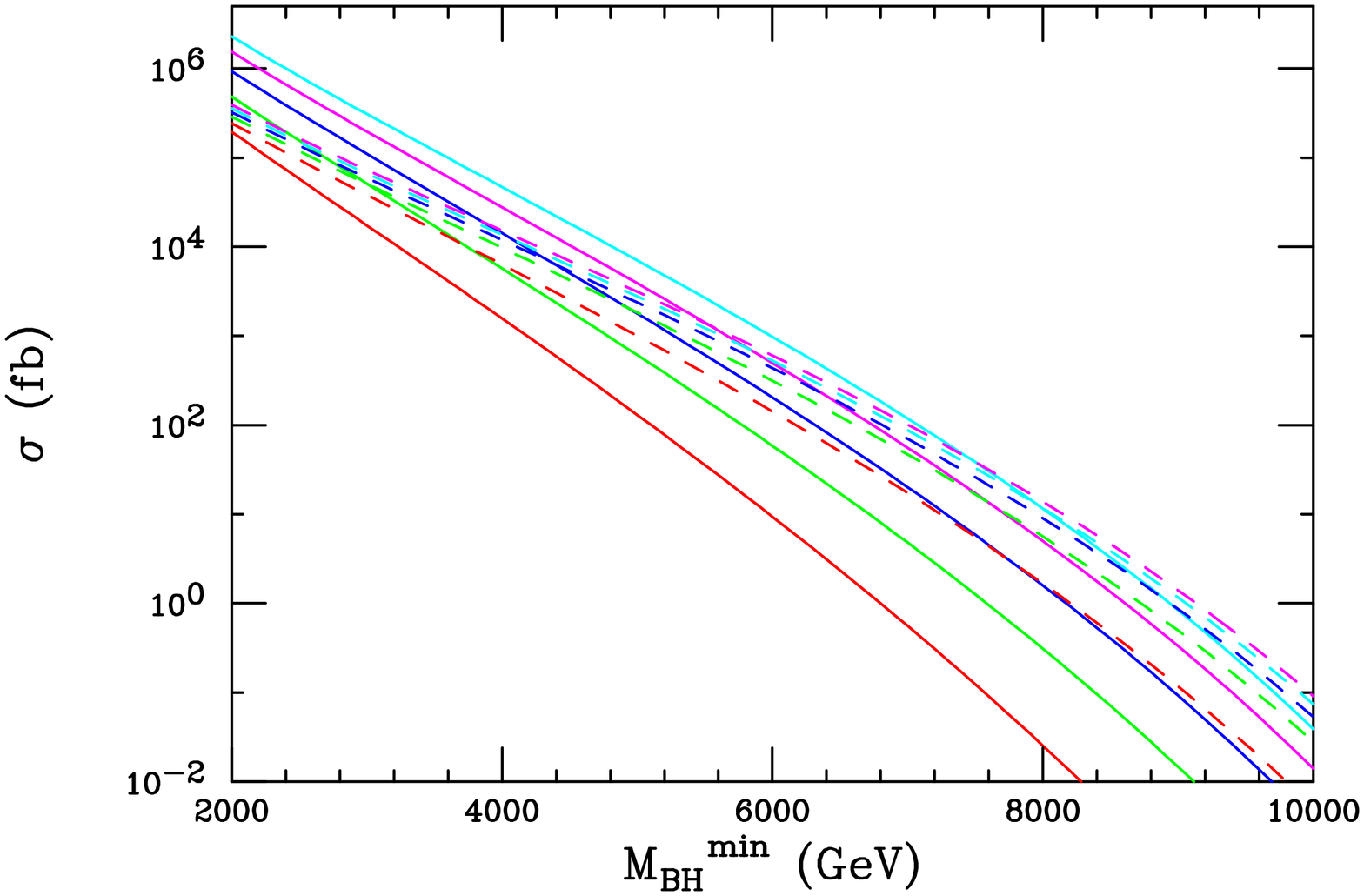}}
\vspace*{0.1cm}
\caption{Same as the last figure but now for a larger value of the 
fundamental Planck scale, $M_* =2$ TeV.}
\label{p3-39_snowholes2}
\end{figure}

Voloshin has recently argued that an additional exponential suppression 
factor, $S$, is also present which seriously damps the 
pure geometric cross section for this process{\cite {voloshin}} even in 
the semi-classical case, \ie, we should rescale 
$\hat \sigma \to S\hat \sigma$ in the equations above. Here $S$ is given by    
\begin{equation}
S=exp[-4\pi R_S M_{BH}/(1+\delta)(2+\delta)] \sim exp \Big[-C
\Big({M_{BH}\over {\mpl}}\Big)^{{2+\delta}\over {1+\delta}}\Big]\,, 
\end{equation}
where $C$ is a relatively small, though $\delta$-dependent, constant. While 
this possibility remains controversial, and strong arguments have 
been made on either side of the argument, for purposes of this discussion we 
will assume this suppression is indeed present. (However, we warn the reader 
that the jury is still out on this issue. In either case we anxiously await 
the resolution of this important argument.) If Voloshin's criticisms of the 
geometrical cross section are valid one worries that the resulting 
exponentially suppressed rates for 
heavy BH production 
will possibly be too small to be observable at the LHC; as we will see below 
this need not be so. 

Just how do the suppressed and unsuppressed cross sections at the LHC compare? 
As can be seen in Fig.~\ref{p3-33_snowholes} for the case $\mpl=1$ TeV, 
the unsuppressed rates for BH production at 
the LHC are quite large over a wide range of masses and numbers of extra 
dimensions using either set of authors' cross section expressions. (In this 
figure and the others below we appropriately 
identify $\mpl$ as either $M_{GT}$ or $M_{DL}$ 
depending on which set of predictions are being discussed.) Note that 
the results of Giddings and Thomas are always larger than those of 
Dimopoulos and Landsberg due to the different definitions used for the Planck 
scale and that the difference between the two sets of 
predictions increases as $\delta$ increases as discussed above. 
We also see that Fig.~\ref{p3-33_snowholes} shows the effects of 
the suppression predicted by Voloshin in the two cases.
From these results we make the important observation that for at least 
for some ranges of parameters BH will still be produced at rates that are 
large enough 
to be observable at the LHC {\it even when the Voloshin suppression is active}. 
For example, assuming that $M_{BH}^{min}=5$ TeV, we see 
that it is quite easy to 
have cross sections in the 100-1000 fb range. Although this is not a huge 
cross section the associated rates at the LHC will be quite large given an 
integrated luminosity of order 100 $fb^{-1}$/yr. 
Note that the suppression factor modifies the two sets of predictions in quite 
different manners due to the two different expressions used for $R_S$. Since 
$(R_S)_{GT}>(R_S)_{DL}$ for all $\delta \geq 2$, assuming the same input 
values for $M_{GT}$ and $M_{DL}$, the GT results are found to be 
more suppressed than are those of DL. 

\begin{figure}[htbp]
\centerline{
\includegraphics[width=5.4cm,angle=90]{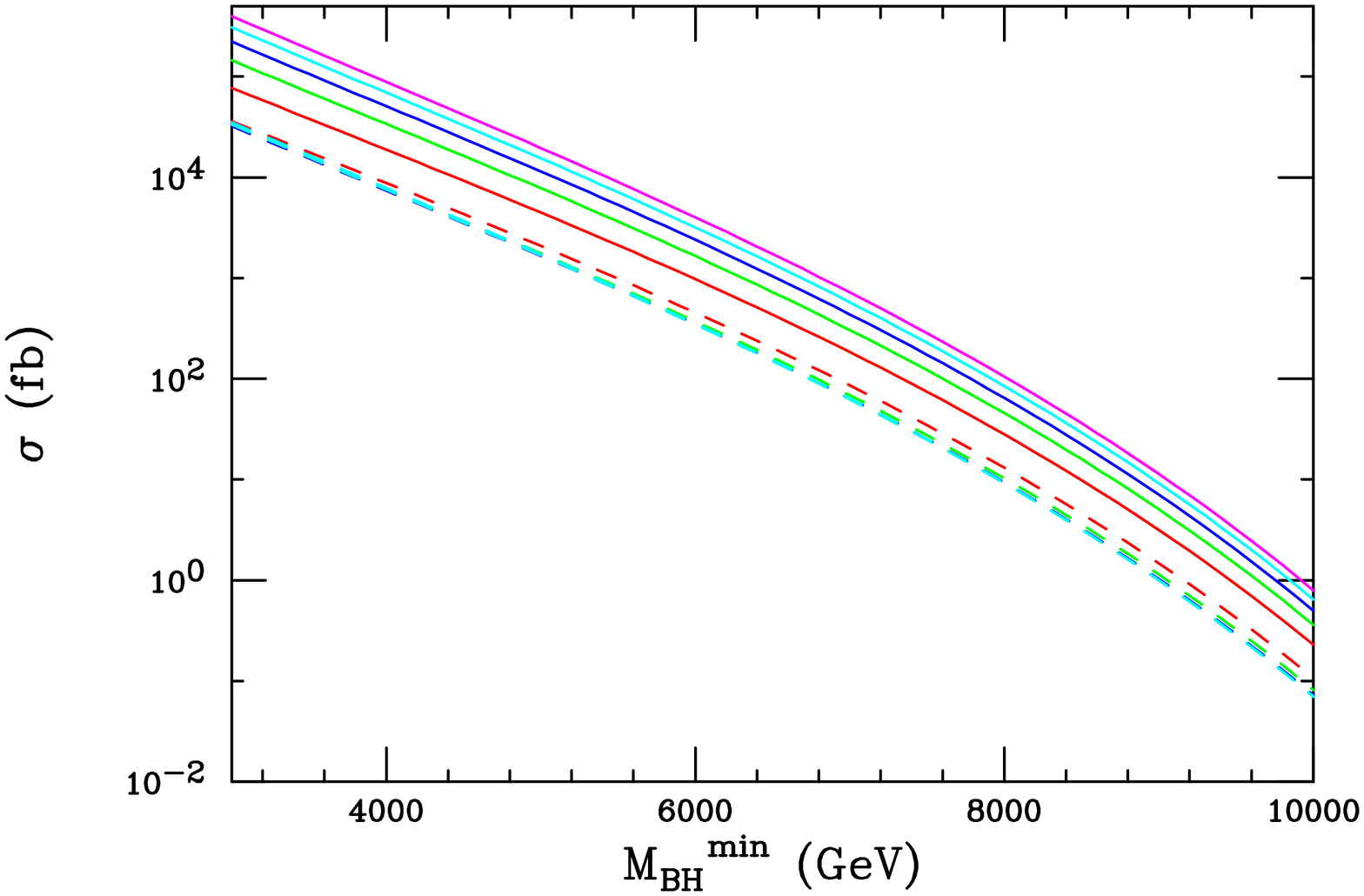}
\hspace*{5mm}
\includegraphics[width=5.4cm,angle=90]{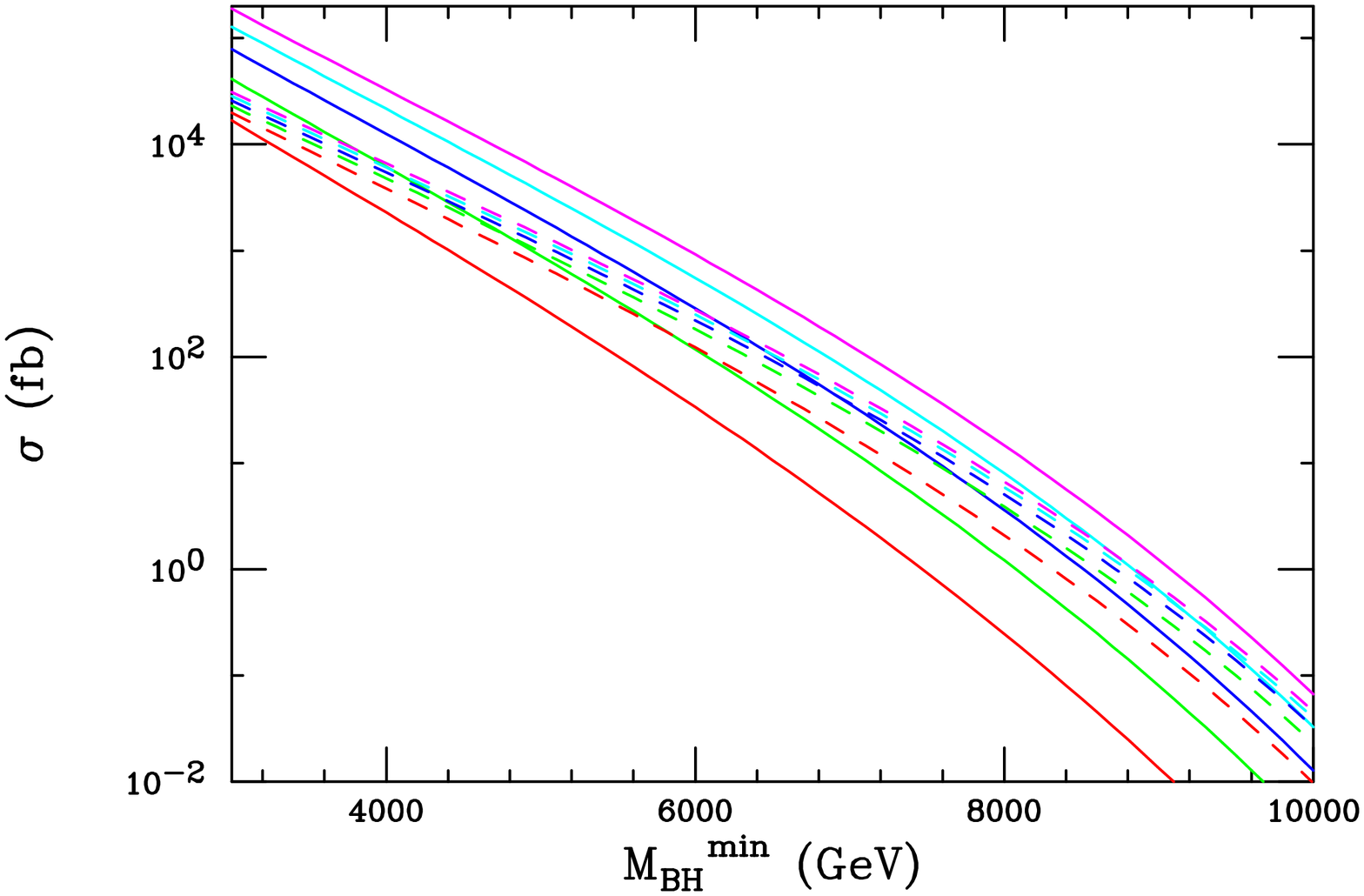}}
\vspace*{0.1cm}
\caption{Same as the last figure but now for $M_* =3$ TeV.}
\label{p3-39_snowholes3}
\end{figure}

What happens as we vary $\mpl$? Figs.~\ref{p3-39_snowholes2} and 
~\ref{p3-39_snowholes3} show the effects of increasing $\mpl$ from 1 TeV to 
2 and 3 TeV. As expected the unsuppressed rates for any fixed value of $M_{BH}$ 
decreases but we also see that the Voloshin suppression becomes less 
effective. This is also to be expected since the ratio $M_{BH}/\mpl$ in the 
exponent of the factor $S$ has been decreased for fixed $M_{BH}$. Again we see 
that for BH in the 5-6 TeV range it is relatively likely that the production 
cross section can quite easily be in excess of 100 fb. 

We remind the reader that once 
produced these BH essentially decay semi-classically, mostly on the brane, 
via Hawking radiation  
into a reasonably large number $\simeq 25$ or more final state partons in a 
highly spherical pattern. Hadrons will dominate over leptons by a factor of 
order 5-10 for such final states. These unusual signatures would not 
be missed at either 
hadron or lepton colliders. (We note that an alternative decay scenario has 
been advocated by Casadio and Harms{\cite {casa}}.) These features are 
sufficiently unique that BH production above conventional backgrounds should 
be observable at the LHC even if the cross sections are substantially smaller 
than the original estimates.

We have examined the production of BH at the LHC assuming that the exponential  
suppression of the geometric cross section predicted by Voloshin is realized.
We have found that even when this suppression is significant the resulting 
rates are still quite large for a wide range of model parameters given an 
integrated luminosity of order 100 $fb^{-1}$.

%
\def\MPL #1 #2 #3 {Mod. Phys. Lett. {\bf#1},\ #2 (#3)}
\def\NPB #1 #2 #3 {Nucl. Phys. {\bf#1},\ #2 (#3)}
\def\PLB #1 #2 #3 {Phys. Lett. {\bf#1},\ #2 (#3)}
\def\PR #1 #2 #3 {Phys. Rep. {\bf#1},\ #2 (#3)}
\def\PRD #1 #2 #3 {Phys. Rev. {\bf#1},\ #2 (#3)}
\def\PRL #1 #2 #3 {Phys. Rev. Lett. {\bf#1},\ #2 (#3)}
\def\RMP #1 #2 #3 {Rev. Mod. Phys. {\bf#1},\ #2 (#3)}
\def\NIM #1 #2 #3 {Nuc. Inst. Meth. {\bf#1},\ #2 (#3)}
\def\ZPC #1 #2 #3 {Z. Phys. {\bf#1},\ #2 (#3)}
\def\EJPC #1 #2 #3 {E. Phys. J. {\bf#1},\ #2 (#3)}
\def\IJMP #1 #2 #3 {Int. J. Mod. Phys. {\bf#1},\ #2 (#3)}
\def\JHEP #1 #2 #3 {J. High En. Phys. {\bf#1},\ #2 (#3)}

\end{document}